%% file: VoiceAI_JMIR.tex
\documentclass[12pt,letterpaper]{article}

\usepackage[utf8]{inputenc}
\usepackage[T1]{fontenc}
\usepackage{times}
\usepackage[margin=1in]{geometry}
\usepackage{setspace}
\usepackage{graphicx}
\usepackage{booktabs}
\usepackage{array}
\usepackage{tabularx}
\usepackage{longtable}
\usepackage{multirow}
\usepackage{amsmath}
\usepackage{hyperref}
\hypersetup{colorlinks=true, linkcolor=blue, citecolor=blue, urlcolor=blue}
\usepackage[numbers,sort&compress]{natbib}
\usepackage{caption}
\usepackage{float}
\usepackage{enumitem}
\usepackage{xcolor}
\usepackage{tikz}
\usetikzlibrary{arrows.meta, positioning, calc}

\title{}
\author{}
\date{}

\begin{document}

\begin{center}
{\Large\bfseries Cross-Domain Transfer of Depression Voice Biomarkers\\
Depends on the Outcome Instrument:\\
Leakage-Controlled Cross-Sectional Evaluation Study}

\bigskip

Rachel L.\ Wiley, MD,$^{1,*,\#}$
James Schwoebel, BSc,$^{2,*}$
Matias Caccia, PhD,$^{3,4}$
Joel Shor, PhD,$^{2,5}$
Adolfo M.\ García, PhD,$^{3,4,6,7}$
Sheehan D.\ Fisher, PhD,$^{8,9}$
Martin G.\ Frasch, MD, PhD$^{8,10,11,\#}$
\end{center}

\bigskip

\noindent
{\small
$^{1}$ Dept.\ of Obstetrics and Gynecology, University of California San Diego, CA, USA\\
$^{2}$ Quome Inc, West Hollywood, CA, USA\\
$^{3}$ TELL Inc, USA\\
$^{4}$ Cognitive Neuroscience Center, University of San Andres, Buenos Aires, Argentina\\
$^{5}$ Move37 Labs LLC, MA, USA\\
$^{6}$ Departamento de Ling\"u\'istica y Literatura, Facultad de Humanidades, Universidad de Santiago de Chile, Santiago, Chile\\
$^{7}$ Global Brain Health Institute (GBHI), University of California, San Francisco, CA, USA\\
$^{8}$ nurtur health, Inc, Derry, NH, USA\\
$^{9}$ Northwestern University, Chicago, IL, USA\\
$^{10}$ Dept.\ of Obstetrics and Gynecology, Institute on Human Development and Disability, University of Washington, Seattle, WA, USA\\
$^{11}$ Health Stream Analytics, LLC, Seattle, WA, USA
}

\medskip
\noindent \medskip

\noindent
$^{*}$ Co-first authorship

\medskip
\noindent $^{\#}$ \textbf{Co-corresponding authors}

\smallskip
\noindent Rachel L.\ Wiley, MD\\
Dept.\ of Obstetrics and Gynecology\\
University of California San Diego, CA, USA\\
Email: rlwiley@health.ucsd.edu 

\smallskip
\noindent Martin G.\ Frasch, MD, PhD\\
Health Stream Analytics, LLC\\
Seattle, WA, USA\\
Email: mfrasch@uw.edu\\
Phone: 206.705.3381

\medskip
\noindent \textbf{Short title:} Outcome-Instrument Dependence of Voice-Biomarker Transfer

\medskip
\noindent \textbf{Word count:} Abstract: $\sim$450 (JMIR: $\leq$450, structured); Main text: $\sim$2,900\\
\textbf{Tables/Figures:} 3 tables, 2 figures (main text); 10 Multimedia Appendices\\
\textbf{References:} $\sim$38 (AMA style, 11th ed.)\\
\textbf{Date:} \today

\noindent
{\small
\textbf{ORCID.}
Rachel L.\ Wiley: \href{https://orcid.org/0000-0003-2158-4482}{0000-0003-2158-4482};
James Schwoebel: \href{https://orcid.org/0000-0002-5697-3894}{0000-0002-5697-3894};
Matias Caccia: \href{https://orcid.org/0000-0002-6770-469X}{0000-0002-6770-469X};
Joel Shor: \href{https://orcid.org/0000-0002-6729-5988}{0000-0002-6729-5988};
Adolfo M.\ García: \href{https://orcid.org/0000-0002-6936-0114}{0000-0002-6936-0114};
Sheehan D.\ Fisher: \href{https://orcid.org/0000-0003-4078-8448}{0000-0003-4078-8448};
Martin G.\ Frasch: \href{https://orcid.org/0000-0003-3159-6321}{0000-0003-3159-6321}.
}

\medskip

\newpage

\setcounter{page}{1}

\begin{center}
\fbox{\begin{minipage}{0.92\textwidth}
\small
\textbf{Changes in this version.} This version supersedes both v1 (May 2025) and
v2 (September 2026), and readers who cited either should treat it as replacing
them in full.

\smallskip
\emph{Relative to v1.} During peer review we identified a scoring defect in the
study's depression labels: the survey platform presented a single generic response
scale in place of the item-specific anchors of the Edinburgh Postnatal Depression
Scale, and the scoring applied to it was non-monotone in symptom severity and did
not reverse-score the two positively-worded items. All analyses were recomputed
under corrected scoring. The correction changed the study's conclusions. The
pre-registered within-cohort outcome is at chance and no cell of the pre-specified
analysis grid survives multiplicity adjustment; an inter-instrument ``ceiling''
argument made in v1 is withdrawn; and the central finding is now that measured
cross-domain transfer varied with the outcome instrument rather than with the
population alone. The analysis was also rebuilt under a leakage-controlled
protocol with participant-grouped cross-validation.

\smallskip
\emph{Relative to v2.} Several reported values are corrected: one confidence
interval in the transfer matrix, the node-coverage figure in the graph-model
appendix, the description of the cohort used to train the forward-transfer model,
and an effect size in the bias appendix that had been taken from a simulation
rather than measured on the study cohort. Multiplicity is now reported under a
single procedure (Benjamini--Hochberg) instead of two. Author order and ORCID
identifiers are corrected, and results under the superseded scoring are no longer
reported alongside the corrected ones.
\end{minipage}}
\end{center}

\bigskip
\section*{Abstract}

\noindent\textbf{Background:} Voice-based depression screening is commercializing rapidly, yet whether voice biomarkers generalize across clinical settings is largely untested. Generalization is usually framed as a question about populations, but it is also a question about the outcome instrument a model is scored against: reported accuracy varies with the diagnostic criteria a study adopts, but confounded with everything else differing between studies.

\medskip
\noindent\textbf{Objective:} We tested whether depression voice biomarkers transfer between perinatal and general-psychiatric speech in both directions under strict leakage control, and whether transfer depends on which self-report instrument defines the outcome.

\medskip
\noindent\textbf{Methods:} In a cross-sectional US-nationwide online study, 446 sessions from 390 pregnant participants at 22$\pm$1 weeks' gestation (analytical sample N=316, one per participant) each provided four voice recordings, the eight-item Patient Health Questionnaire (PHQ-8), and a modified nine-item Edinburgh Postnatal Depression Scale (mEPDS-9). We extracted 375 acoustic descriptors and 330-dimensional Population Dynamics Foundation Model (PDFM) community-health embeddings. Discrimination (area under the curve, AUC) was assessed under leakage-controlled cross-validation, with feature and classifier selection inside training folds only, gated by a permutation negative-control harness, across a pre-specified grid of 4 speech tasks $\times$ 4 outcome definitions. The prenatal-trained model was applied to two general-psychiatric corpora (DAIC-WOZ, N=189; E-DAIC, N=219) under pure held-out inference, and an open-source model (DAM) trained on $\sim$35{,}000 individuals was applied to all three cohorts without refitting. Multiplicity was controlled by the Benjamini--Hochberg procedure.

\medskip
\noindent\textbf{Results:} The pre-registered within-cohort outcome performed at chance (AUC 0.494, 95\% CI 0.431--0.560), and no cell of the 16-cell grid survived multiplicity adjustment under either modeling paradigm. The prenatal-trained model did not transfer to either general-psychiatric corpus (AUC 0.505, 0.478). Transfer measured in the reverse direction, however, varied with the outcome instrument: DAM reached AUC 0.706--0.708 on general-psychiatric speech, 0.510 (0.411--0.609) against the PHQ-8 in pregnant women, and 0.645 (0.541--0.744) against the mEPDS-9 \emph{in the same women}; paired difference 0.135 (0.019--0.248; unadjusted post-hoc $p$=0.021). An exploratory item-level analysis suggests a mechanism: DAM tracked affective-distress items (crying $r$=0.26, sadness $r$=0.21, overwhelmed $r$=0.20) but not the somatic PHQ-8 items, though only crying survived adjustment across 17 tests. Transfer also varied descriptively by elicitation task (pregnancy narrative 0.653, 95\% CI 0.550--0.753; open-ended speech 0.597, 0.486--0.699; scripted reading 0.544, 0.434--0.651), not compared formally.

\medskip
\noindent\textbf{Conclusions:} Whether a depression voice biomarker transfers may depend not only on the population but on whether the outcome instrument captures what the model can detect: in the same pregnant participants, a general-population speech model discriminated better against the mEPDS-9 than against the PHQ-8. Validation of voice-based mental-health tools should state the population, the elicitation task, and the outcome instrument together, and a tool validated against one instrument should not be assumed to work against another.

\medskip
\noindent\textbf{Trial Registration:} Preregistered on the Open Science Framework.\cite{schwoebel2025preprint}

\medskip
\noindent\textbf{Keywords:} perinatal depression; voice biomarkers; digital phenotyping; speech analysis; machine learning; data leakage; external validation; model generalizability; depression screening; Population Dynamics Foundation Model

\newpage

\section*{Introduction}

Perinatal depression affects up to 20\% of pregnant individuals and is a leading cause of maternal morbidity and pregnancy-related death in the United States, yet only about half of antenatal cases are identified during pregnancy and fewer than 10\% receive adequate treatment.\cite{bauman2020,cox2016,trost2021} Postpartum-depression prevalence roughly doubled over the past decade in large US cohorts (9.4\% in 2010 to 19.0\% in 2021 in a Kaiser Permanente Southern California cohort of 442{,}308 pregnancies), with the steepest increases among Asian/Pacific-Islander and non-Hispanic Black participants.\cite{khadka2024,logue2021} Current screening commonly relies on standardized self-report questionnaires such as the EPDS and PHQ-9, administered or reviewed within clinical care and recommended during perinatal care,\cite{acog2023} but real-world adherence is variable and even adherent screening can miss interval mood changes. Scalable, low-burden adjuncts that integrate into existing workflows are needed.

Voice carries a biological signal of depressive state: major depression produces psychomotor retardation and altered laryngeal tone, manifesting as reduced fundamental-frequency variability, longer pauses, reduced speech rate, and elevated jitter/shimmer.\cite{low2020,koops2023,kappen2023} Pregnancy itself produces overlapping but distinct phonatory changes via progesterone-mediated mucosal effects and altered respiratory mechanics.\cite{hancock2014,ghaemi2018,rechenberg2022} This overlay motivates the usual framing of the question: are depression voice biomarkers population-specific: do models trained on general-population speech generalize to perinatal speech, where the phonatory baseline is itself perturbed? We test that question in both directions, and add one the literature has left confounded. A systematic review of 25 speech-based depression studies reports that diagnostic performance varied with the diagnostic criteria used, alongside sample size, validation strategy and language.\cite{lu2025speechmeta} Because those criteria differ between studies at the same time as the cohorts, the recordings and the models do, the instrument's contribution cannot be separated from the rest; we are not aware of a test that holds participants, recordings and model fixed and varies only the instrument. Every such model is scored against a self-report instrument, and instruments differ in what they measure; in pregnancy in particular, somatic-weighted instruments capture symptoms that gestation elevates independently of mood. Generalization may therefore depend not only on who is speaking but on what the outcome measure is asking. Our cohort carries two instruments on the same participants, which allows population and instrument to be separated in one direction.

We address this and a second, methodological question. This cohort was recruited cross-sectionally at the single mid-gestation timepoint of 22$\pm$1 weeks, so the findings concern antenatal mid-gestation depression specifically. To enrich acoustic screening without added patient burden, we incorporated community-level health embeddings from Google's Population Dynamics Foundation Model (PDFM),\cite{google_pdfm} which summarizes socioeconomic conditions, healthcare access, and community-health patterns for US ZIP codes already captured in medical records.

We tested generalizability against two independent general-psychiatric datasets (DAIC-WOZ, E-DAIC) and an open-source deep-learning model trained on $\sim$35{,}000 individuals.\cite{kintsugi_dam} We had two hypotheses. First, that depression voice biomarkers would prove population-specific, transferring poorly between prenatal and general-psychiatric contexts. Second, that within a prenatal cohort, acoustic features fused with PDFM embeddings would yield a signal strong enough to be deployable. Both were tested under a strict leakage-control protocol: an \texttt{sklearn} \texttt{Pipeline} fitted inside cross-validation, no pre-split class balancing, no operating-point selection on reported predictions, and a 20-permutation negative-control harness. We offer that protocol as a methodological template.

\section*{Methods}

\subsection*{Study Design and Participants}

This was a cross-sectional US-nationwide online study of pregnant individuals at 22$\pm$1 weeks' gestation. The study was pre-registered on the Open Science Framework (OSF).\cite{schwoebel2025preprint} Participants were recruited via Cint, an online survey-recruitment platform that maintains a verified panel of US-based participants,\cite{cint} and directed to the secure Databoard platform to consent and complete the study. The study was deemed ``not human subjects research'' by the University of California San Diego Institutional Review Board (IRB\#812757) as it collected no identifiable information.

Participants were included if female, aged 18--45, pregnant at 22$\pm$1 weeks, able to respond via a computing device with internet access, and without current suicidal ideation.

\input{figure_consort}

\subsection*{Voice Recording Protocol}

Participants completed four vocal tasks via internet browser: (1) a 10-second sound check, (2) a 1-minute free speech prompt on pregnancy experiences, (3) a 2-minute structured reading passage, and (4) a 1-minute free speech prompt on a non-pregnancy topic. Tasks 2 and 4 were counterbalanced across participants to control for order effects (Multimedia Appendix 2).

\subsection*{Depression Assessment}

Depression was measured at the single 22$\pm$1-week timepoint using the eight-item Patient Health Questionnaire (PHQ-8) and the Edinburgh Postnatal Depression Scale\cite{cox1987epds} with item 10 (suicidality) removed (IRB-mandated).\cite{kroenke2002,kroenke2009} Binary classification used $\geq$10 and $\geq$13 thresholds for both instruments. \textbf{The pre-registered primary outcome is Reading Passage with PHQ-8$\geq$10}; all mEPDS-9 analyses are secondary or exploratory, and all four thresholds are reported.

A caution about the mEPDS-9 thresholds is required. The instrument as administered used a generic four-point agreement scale, not the item-specific response anchors of the published EPDS, and omitted item 10. Correcting the scoring restores a monotone symptom score, but it does not establish that the psychometric properties of the published EPDS (including the validation of its $\geq$10 and $\geq$13 cutoffs) transfer to a nine-item instrument administered with different response options. We therefore treat mEPDS-9 scores as a reconstructed symptom score, and the $\geq$10 and $\geq$13 values as operational thresholds mapped from EPDS convention, not as validated diagnostic cutoffs for this administration.

\subsection*{Rationale for a Single-Timepoint Screening Outcome}

The study evaluates voice analysis as a \emph{screening} adjunct, and its real-world comparator is exactly what was administered here: a single self-report screening instrument at a single clinical touchpoint, which is how perinatal depression screening is delivered in practice.\cite{acog2023} A screen-positive result is not a diagnosis. Individual-participant-data meta-analyses report that against semi-structured diagnostic interview the PHQ-9 at $\geq$10 achieves sensitivity 0.85 (95\% CI 0.79--0.89) and specificity 0.85 (0.82--0.87),\cite{negeri2021phq9ipdma} and the EPDS at $\geq$10 sensitivity 0.85 (0.79--0.90) and specificity 0.84 (0.79--0.88).\cite{levis2020epdsipdma} That is the performance level a screening adjunct would have to approach to be useful, and it is far above anything observed here.

Two caveats attach to the EPDS figure. First, the same meta-analysis found that combined sensitivity and specificity is maximised not at $\geq$10 but at $\geq$11; we use $\geq$10 because it is the conventional threshold and the one specified in our pre-registration, not because it is the optimal operating point in that analysis. Second, and more importantly, these figures apply to the published instruments as validated, and cannot be assumed to transfer to the modified administration used here (see below). Throughout, the outcome is described as screen-positive depressive symptomatology rather than depression as a diagnostic entity.

Both instruments were self-administered online, unassisted, in the same session as the voice recordings; no clinician was involved in administration or scoring. The platform presented, for all nine mEPDS-9 items, a single generic four-point agreement scale (``Yes, all of the time'' / ``Yes, most of the time'' / ``No, not very often'' / ``No, not at all'') instead of the item-specific anchors of the published EPDS.

\subsection*{Acoustic Feature Extraction (the ``privacy tier'')}

We extracted 375 audio-only acoustic descriptors per recording, composed of four feature blocks. Block sizes and the open-source toolboxes used are stated explicitly:

The 375 descriptors comprise four blocks: a 325-feature base-acoustic block (13 MFCCs, 26 log-Mel-filterbank energies, 26 spectral subband centroids, each summarized by five statistics) adapted from the open-source Allie ML framework's tier-1 audio pipeline;\cite{schwoebel_allie,schwoebel2021longitudinal} 22 original voice-quality features (F0, jitter, shimmer, HNR, formants) via \texttt{parselmouth}/Praat;\cite{boersma_praat,parselmouth} 12 extended voice-quality distributional moments; and 16 talking-interval/timing features. Per-block extraction detail is in Multimedia Appendix 3.

These features measure \textit{how} speech is produced; spectral shape, voice-source characteristics, and timing, and require \emph{no speech transcription, no language model, and no semantic content}. We refer to this block collectively as the ``privacy tier'' throughout the main-text Results, the calibration grid, the negative-control harness, and the leakage-controlled TPOT-CV sensitivity analysis (Multimedia Appendix 8). Audio loading uses \texttt{soundfile} via a thin wrapper (\texttt{pipeline\_a/audio\_loader.py}).

Within each cross-validation training fold, the 375-feature vector was reduced to 20 features by univariate filter feature selection (using \texttt{SelectKBest} with the ANOVA F-statistic from \texttt{scikit-learn}), with no information from the held-out test fold used in the selection. Technical details of the per-block extraction pipelines, together with documentation of two additional feature sources (TELL linguistic features and Kintsugi DAM scores) that are deliberately excluded from the headline analysis, are in Multimedia Appendix 3.

\subsection*{Community Health Embeddings}

PDFM provides 330-dimensional embeddings per US ZIP Code Tabulation Area (ZCTA), summarizing community-level socioeconomic conditions, healthcare access, environmental factors, and health patterns from publicly available data.\cite{google_pdfm} We merged pre-trained embeddings for 32{,}069 US ZCTAs with participant data; 88.9\% of participants had a matching ZCTA embedding, and unmatched participants were excluded from the headline analytical sample (see Figure~\ref{fig:consort}). As with the acoustic features, the 330-dimensional embedding was reduced to 20 features by \texttt{SelectKBest} fit on the training fold only.

Critically, PDFM uses only the ZIP code already present in medical records. It requires no additional demographic questionnaires, no individual-level data collection, and no patient interaction; making it a zero-burden enhancement applicable in any clinical setting with access to a patient's address.

We retained ZCTA-level instead of coarser (e.g.\ urban/rural) granularity; a parallel large-cohort PRAMS analysis indicates binary urban/rural stratification of PDFM underperforms finer schemes, so coarse aggregation discards signal.\cite{frasch2026prams}

\subsection*{Model Development under Leakage-Controlled Cross-Validation}

Acoustic features and PDFM embeddings were combined by late fusion: a separate \texttt{StandardScaler} $\rightarrow$ \texttt{SelectKBest} $\rightarrow$ classifier pipeline was refit per cross-validation fold on each modality's features, and the two arms' out-of-fold positive-class probabilities were averaged with equal weight to form the fused score (no meta-learner, no fitted fusion weights). The analysis uses a strictly leakage-free protocol (Multimedia Appendix 1): an \texttt{sklearn} \texttt{Pipeline} containing \texttt{StandardScaler} $\rightarrow$ \texttt{SelectKBest} $\rightarrow$ classifier was passed to \texttt{cross\_val\_predict} with 5-fold stratified cross-validation, ensuring scaling and feature selection are refit on each training fold. Class imbalance was handled by setting the classifier's \texttt{class\_weight} parameter to \texttt{`balanced'}; no pre-split downsampling was applied. Three evaluation protocols are reported. (a) A parsimonious single-classifier model using L2-regularized logistic regression, which provides the headline estimate. (b) A sensitivity analysis using nested cross-validation that selects between logistic-regression, gradient-boosting, and random-forest candidates inside an inner cross-validation loop on training data only. (c) An automated machine-learning (AutoML) sensitivity analysis using TPOT,\cite{olson2016tpot} in which the pipeline itself; preprocessing, feature transformation and classifier; is searched rather than fixed in advance. In (c), outer folds are participant-grouped (\texttt{StratifiedGroupKFold}), \texttt{StandardScaler} and \texttt{SelectKBest}(k=20) are fitted on outer-training rows only, and TPOT runs its own internal cross-validation on those rows to select a pipeline, leaving the outer-test partition untouched until prediction; the search budget was three minutes per outer fold with the \texttt{linear-light} search space. Protocols (b) and (c) are the two modeling paradigms across which the pre-specified grid is reported; the full AutoML specification is in Multimedia Appendix 8. Because the analytical sample retains at most one session per participant (dual-session-artifact participants were removed entirely; Figure~\ref{fig:consort}), cross-validation folds contain no within-participant repeated measures; with one session per participant, grouping by participant identifier (\texttt{GroupKFold}) is equivalent to the stratified folds used here, so no participant-level leakage is possible.

Operating-point metrics used the fixed 0.5 threshold and a sensitivity-targeted threshold. \emph{No threshold was tuned to improve reported performance.} The sensitivity-targeted threshold was selected inside training folds only: the threshold reaching training-fold sensitivity $\geq$80\% was chosen on training rows and applied unchanged to held-out rows. This is a leakage control, not an optimisation step; choosing an operating point on the same predictions that are then reported would inflate every metric derived from it. Discrimination (AUC) is threshold-independent in any case, so the reported AUCs do not depend on either choice. PPV/NPV were Bayes-recalibrated to 12--30\% prevalence (Table~\ref{tab:screening}). Bootstrap 95\% confidence intervals (1{,}000 resamples) were computed for all AUC estimates, with pairwise contrasts by DeLong's test.\cite{delong1988}

\subsection*{Negative-Control Harness}

Before reading the AUC for any pipeline we ran a 20-permutation negative-control harness (shuffled-label, random-feature, and joint-shuffle controls); a clean pipeline must produce AUC $\approx$0.5 in all three conditions, and all pre-registered and exploratory configurations passed. Full harness specification is in Multimedia Appendix 1.

\subsection*{External Validation}

We evaluated prenatal-trained models on two independent general-psychiatric datasets: DAIC-WOZ ($N=189$) and E-DAIC ($N=219$), both semi-structured clinical interviews of mixed-gender adults, and benchmarked against the Kintsugi Depression/Anxiety Model (DAM), an open-source model fine-tuned from Whisper on $\sim$863 hours of calls from $\sim$35{,}000 individuals (Multimedia Appendix 4).\cite{gratch2014,ringeval2017,kintsugi_dam} All cross-domain AUCs reflect pure inference: no refitting, recalibration or threshold selection was performed on the target cohorts, and cross-domain operating points use the fixed 0.5 threshold. For the prenatal$\to$external direction, the privacy-tier acoustic block with a leakage-controlled logistic-regression pipeline (StandardScaler $\rightarrow$ SelectKBest(20) $\rightarrow$ class-weighted L2 LR) was fitted on the Reading-Passage rows of the analytical sample carrying PHQ-8 labels (N=312; Figure~\ref{fig:consort}) and applied to native DAIC-WOZ and E-DAIC features extracted by the identical pipeline (375/375 feature alignment); PDFM is unavailable externally, so the acoustic arm is used (implemented in the external-validation scripts; e.g.\ \texttt{run\_prenatal\_to\_edaic.py}).

\subsection*{Statistical Analysis}

Data were tested for normality with appropriate two-tailed group comparisons at $\alpha=0.05$; demographic confounding was assessed with effect sizes and inverse probability weighting (Multimedia Appendix 5).

\textit{Multiplicity.} Two families of tests are adjusted for multiple comparisons: the pre-specified 16-cell task $\times$ outcome grid, and the comparator-model cells evaluated on the prenatal cohort. Within each family, one-sided $p$ values for H$_0$: AUC $=0.5$ were obtained from the Mann--Whitney statistic on out-of-fold predictions and adjusted by the Benjamini--Hochberg procedure.\cite{benjamini1995} We report one adjustment procedure, not several. Benjamini--Hochberg controls the false discovery rate instead of the family-wise error rate and is the less conservative of the two procedures in common use; at this sample size a family-wise correction would have little power against effects of the size at issue, and reporting both procedures would invite a reader to adopt whichever is convenient once the result is known. The two families are adjusted separately because they address distinct analytic questions. Pipeline scripts, the negative-control harness, and per-(task$\times$threshold) result JSONs are available on reasonable request (see Data Availability).

\subsection*{Ethical Considerations}

The University of California San Diego Institutional Review Board determined that the study did not constitute human-subjects research (determination IRB\#812757). We state the determination without restating the rationale for it; we note that ZIP code, which this study uses, can be identifying in combination with other attributes. Participants reviewed and agreed to a digital consent statement on the secure Databoard platform before any data collection.

Current suicidal ideation was an exclusion criterion applied by the recruitment panel at eligibility screening, before enrolment and before any study instrument was administered; the study team did not administer a suicidality item at any point, and the suicidality item (EPDS item 10) was omitted from both administered instruments per the IRB determination. Because no suicidality item was administered within the study, no in-study endorsement could occur and collection, and a deployed system would require a separate suicidality-management pathway. No compensation details, identifiable demographics, or contact information were retained, and the terms of consent did not authorize sharing of the prenatal voice data.

\section*{Results}

\subsection*{Participants}

The full data assembly comprised 446 sessions from 390 unique participants. Two participant-grouping artifacts defined the analytical sample: 31 participants contributed duplicate sessions submitted 6--24 minutes apart (often with scores differing by 10+ points and crossing the $\geq$10 cutoff in opposite directions), treated as platform artifacts and excluded; and 16 sessions missing protocol-specified audio were excluded. The analytical sample is $N=316$ (Figure~\ref{fig:consort}). Participants were aged $\sim$30$\pm$7 years with no age difference between groups. Cohort prevalence was elevated relative to general perinatal populations (PHQ-8$\geq$10 55.8\%, modified-EPDS-9$\geq$10 73.5\%), reflecting online-recruitment selection toward depressive symptomatology. Both instruments were administered in IRB-approved truncated form (suicidality item omitted). Both instruments show good internal consistency (Cronbach's $\alpha=0.882$ for the PHQ-8 and $0.881$ for the mEPDS-9; McDonald's $\omega=0.883$ and $0.879$) and agree with one another at Pearson $r=0.660$ (Spearman $\rho=0.639$; Cohen's $\kappa$ at $\geq$10 $=0.41$), comparable to the median PHQ-9--EPDS correlation of $r=0.59$ reported across perinatal validation studies.\cite{wang2021phq9perinatal} One instrument's continuous score predicts the other's $\geq$10 binary at AUC $0.80$--$0.84$.

\subsection*{Within-Cohort Screening Performance}

Under the leakage-controlled protocol on the analytical sample, the pre-registered outcome; PHQ-8$\geq$10 on the Reading Passage task, privacy-tier acoustic features fused with PDFM: performed at chance: AUC 0.494 (95\% CI 0.431--0.560). The acoustic arm alone reached 0.452 and the PDFM arm alone 0.541, so the voice increment over community context alone was negative ($\Delta\text{AUC}=-0.047$). At the fixed 0.5 threshold, sensitivity was 0.64 and specificity 0.38, giving LR$+$ $=1.02$ and LR$-$ $=0.96$. Likelihood ratios this close to 1 mean a positive or negative result leaves a patient's estimated probability of depression essentially where it started, and Bayes-recalibrated PPV accordingly approximates prevalence at every deployment prevalence examined (Table~\ref{tab:screening}). The same cell under the AutoML pipeline-search protocol was also at chance (AUC 0.540, 95\% CI 0.477--0.603), so the null does not depend on the choice of classifier or on fixing the pipeline in advance. The configuration passed the permutation negative-control harness (real 0.505, shuffled labels 0.508, random features 0.504), confirming that the null reflects absent signal and not a broken pipeline. The harness applies a single fixed logistic-regression pipeline to the concatenated feature matrix, and not the nested classifier-selection and late-fusion procedure used for the reported estimate, which is why its ``real'' value (0.505) differs slightly from the headline AUC (0.494); the comparison of interest is between the real run and its own shuffled controls, not against the headline.

\begin{table}[H]
\centering
\caption{Operating-point metrics for the pre-registered configuration, with deployment-prevalence Bayes recalibration. LR$+$ $=1.02$ and LR$-$ $=0.96$; essentially non-informative, so PPV approximates prevalence and NPV approximates 1$-$prevalence at every deployment prevalence, consistent with an AUC of 0.494. Sens, Spec, LR$+$ and LR$-$ are prevalence-independent; PPV, NPV and NNS are recomputed by Bayes' theorem across the 12--30\% deployment range (first row shows in-cohort prevalence). Row format is task / feature set / outcome. Abbreviations: RP, reading passage; privacy + PDFM, privacy-tier acoustic features fused with Population Dynamics Foundation Model community-health embeddings; Prev, prevalence; Sens, sensitivity; Spec, specificity; PPV/NPV, positive/negative predictive value; NNS, number needed to screen; LR$+$/LR$-$, positive/negative likelihood ratio.}
\label{tab:screening}
\setlength{\tabcolsep}{3.5pt}\small
\begin{tabular}{lcccccccc}
\toprule
\textbf{Config} & \textbf{Prev} & \textbf{Sens} & \textbf{Spec} & \textbf{PPV} & \textbf{NPV} & \textbf{NNS} & \textbf{LR+} & \textbf{LR$-$} \\
\midrule
RP / privacy + PDFM / PHQ-8$\geq$10 & 55.8\% & 0.638 & 0.377 & 0.563 & 0.452 & 1.8 & 1.02 & 0.96 \\
RP / privacy + PDFM / PHQ-8$\geq$10 & 12\% & 0.638 & 0.377 & 0.122 & 0.884 & 8.2 & 1.02 & 0.96 \\
RP / privacy + PDFM / PHQ-8$\geq$10 & 15\% & 0.638 & 0.377 & 0.153 & 0.855 & 6.5 & 1.02 & 0.96 \\
RP / privacy + PDFM / PHQ-8$\geq$10 & 20\% & 0.638 & 0.377 & 0.204 & 0.806 & 4.9 & 1.02 & 0.96 \\
RP / privacy + PDFM / PHQ-8$\geq$10 & 25\% & 0.638 & 0.377 & 0.254 & 0.757 & 3.9 & 1.02 & 0.96 \\
RP / privacy + PDFM / PHQ-8$\geq$10 & 30\% & 0.638 & 0.377 & 0.305 & 0.708 & 3.3 & 1.02 & 0.96 \\
\bottomrule
\end{tabular}
\end{table}

\textit{The full pre-specified grid, with multiplicity adjustment.} Across all 16 task $\times$ outcome cells, no cell reached nominal significance under nested cross-validation (maximum AUC 0.550); one cell reached the nominal level under AutoML pipeline search (maximum AUC 0.569, raw $p=0.020$) but did not survive adjustment. Every confidence interval covered 0.5 under nested cross-validation. After Benjamini--Hochberg adjustment across the grid, \emph{no cell survives under either modeling paradigm} (Table~\ref{tab:multiplicity}). Under a global null the expected number of cells clearing the nominal 5\% level is 0.8, which is what we observe.

\begin{table}[H]
\centering
\caption{Multiplicity across the pre-specified 16-cell grid (4 speech tasks $\times$ 4 outcome definitions), under two modeling paradigms. $p$ is one-sided for H$_0$: AUC $=0.5$, from the Mann--Whitney statistic on out-of-fold predictions; adjustment is by the Benjamini--Hochberg procedure across the 16 cells within each row. Under a global null the expected number of cells clearing the nominal 5\% level is 0.8. Abbreviation: BH, Benjamini--Hochberg.}
\label{tab:multiplicity}
\setlength{\tabcolsep}{3.5pt}\small
\begin{tabular}{lccc}
\toprule
\textbf{Paradigm} & \textbf{Max AUC} & \textbf{Cells $p<0.05$} & \textbf{Surviving BH} \\
\midrule
AutoML pipeline search & 0.569 & 1 & 0 \\
Nested cross-validation & 0.550 & 0 & 0 \\
\bottomrule
\end{tabular}
\end{table}

\textit{Cell-level estimates in this grid are unstable.} Repeating the AutoML search under protocol variations reproduces individual cell values only loosely: mean absolute difference 0.036, maximum 0.081, and correlation across the 16 cells $r=0.376$. The search is stochastic under a wall-clock budget, and cell values additionally shift with the fold-grouping scheme and with whether item-complete subsets are used, so no single cell of this grid should be interpreted on its own. This is why the grid is reported in full with multiplicity adjustment rather than as a single best-performing configuration; selecting a primary outcome by choosing the highest cell would be reading noise.

\textit{Sensitivity analyses.} A reduced 325-feature acoustic tier fused with PDFM performed no better (AUC 0.501 at the pre-registered outcome). Removing the participant-grouping filter did not change the conclusion (N=372, AUC 0.540). On the subset with acoustic, DAM and PDFM features jointly available (N=115), no block among seven combinations reached significance (maximum 0.576), and adding DAM to voice$+$PDFM produced no increment in any cell (largest $\Delta\text{AUC}=+0.034$, 95\% CI $-0.012$ to $+0.083$). Adding the individually collected sociodemographic variables (age, education, employment, race, household income, relationship status and smoking) did not improve performance ($\Delta\text{AUC}=+0.009$ over voice $+$ PDFM; the sociodemographic block alone reached 0.508), nor did semantic transcript features.

\subsection*{Transfer Depends on the Outcome Instrument, Not Only the Population}

Cross-domain evaluation is a \emph{compound} shift: population, elicitation task and outcome instrument all differ at once, and a design that varies them together cannot say which one matters. In one direction, however, we could hold population and task fixed and vary only the instrument, because our cohort carries two outcome measures on the same participants. That comparison is the central result of this study.

\textit{Prenatal $\rightarrow$ general-psychiatric population: no transfer.} Under pure held-out inference (prenatal-trained leakage-controlled acoustic model; no refitting on target), the model did not separate from chance on either corpus: DAIC-WOZ AUC 0.505 [0.417, 0.593], E-DAIC 0.478 [0.398, 0.564]. This held across feature-extraction toolkits: our features, OpenSMILE eGeMAPS (0.483) and OpenSMILE MFCCs (0.546) were all near chance on E-DAIC (Multimedia Appendix 6), so it is not a toolkit artifact.

\begin{figure}[H]
\centering
\includegraphics[width=0.8\textwidth]{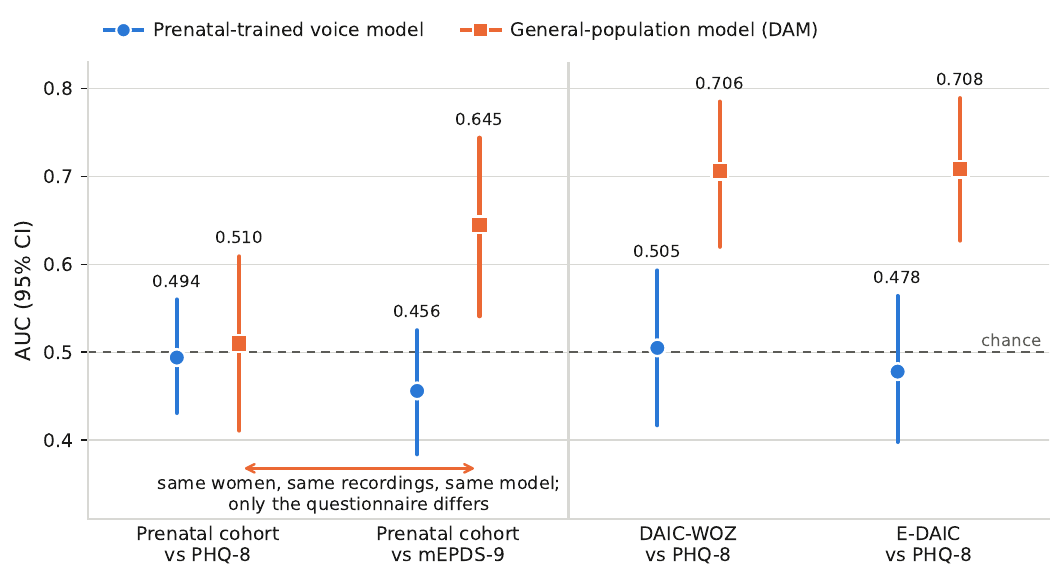}
\caption{Transfer of depression voice biomarkers across cohorts and outcome instruments. Points are AUC; bars are 95\% bootstrap confidence intervals; the dashed line marks chance. All cross-domain estimates are pure held-out inference; the prenatal-trained model's two prenatal conditions are leakage-controlled cross-validated estimates within its own cohort. The prenatal-trained model (blue circles) does not separate from chance in any condition, including on its own cohort. The general-population model (DAM, orange squares) is competent on both general-psychiatric corpora (0.706--0.708), at chance against the PHQ-8 in pregnant women (0.510), and above chance against the modified EPDS-9 in the same women (0.645). The two left-hand conditions share participants, recordings and model, and differ only in the questionnaire defining the outcome; that contrast is the study's central finding. As in Table~\ref{tab:crossdomain}, the two prenatal conditions for the prenatal-trained model use the acoustic + community-context fusion while its two external conditions use the acoustic block alone, because community-health embeddings do not exist for DAIC-WOZ or E-DAIC; the comparator model is acoustic-only in all four. Confidence intervals are shown instead of a colour-coded matrix because several of these estimates are not distinguishable from one another, and a shaded matrix would imply a precision the data do not support. Series are distinguished by hue, marker shape and direct labels, so the figure remains readable in greyscale.}
\label{fig:crossdomain}
\end{figure}

\textit{General-psychiatric $\rightarrow$ prenatal population: measured transfer differed substantially by outcome instrument.} DAM, trained on $\sim$35{,}000 individuals and applied without refitting, reached AUC 0.706--0.708 on the two general-psychiatric corpora. Applied to pregnant women it reached 0.510 [0.411, 0.609] against the PHQ-8 and 0.645 [0.541, 0.744] against the mEPDS-9: \emph{in the same participants, on the same recordings, differing only in which questionnaire defined the outcome} (Table~\ref{tab:crossdomain}, Figure~\ref{fig:crossdomain}). \textbf{The two estimates differ from one another on a direct paired test}, which the design permits because both are computed on the same participants and recordings: paired bootstrap $\Delta$AUC $=+0.135$ (95\% CI $+0.019$ to $+0.248$; unadjusted $p=0.021$; 10{,}000 resamples, participants resampled once per iteration and both AUCs recomputed on that draw so their correlation is preserved). The contrast is significant on all three cohort definitions ($+0.138$, $p=0.011$ at N=149; $+0.127$, $p=0.037$ at N=115). This contrast was not pre-specified, is not part of the 16-cell multiplicity family reported above, and the $p$ value is unadjusted; it requires independent replication. The pre-specified 16-cell task $\times$ outcome grid and this comparator-model family are treated as separate multiplicity families because they address distinct analytic questions. The mEPDS-9 estimate survives Benjamini--Hochberg adjustment across the 10 comparator-model-on-prenatal cells examined (four outcome definitions plus per-task breakdowns) at adjusted $p=0.005$, and is stable across cohort definitions: 0.651 [0.549, 0.746] on all N=149 participants with DAM-processable audio, 0.645 [0.541, 0.744] after applying the same participant-grouping exclusions used for the main analysis (N=136, reported here), and 0.642 [0.528, 0.749] on the N=115 who additionally meet the reading-passage and ZIP-matching requirements of the main analytical sample. The PHQ-8 estimate is 0.510--0.515 in all three.

A general-population speech model therefore showed evidence of transfer into pregnancy when evaluated against the mEPDS-9, but not when evaluated against the PHQ-8.

\textit{Exploratory evidence consistent with construct mismatch.} The analyses in this paragraph were not pre-specified, are based on correlations of about 0.2 in a subset of 136 participants, and are reported as hypothesis-generating. Across the full family of 17 items examined (nine mEPDS-9, eight PHQ-8), one item survives Benjamini--Hochberg adjustment (crying, $r=0.26$, adjusted $p=0.025$); the remaining associations are nominal only. The pattern is offered as a candidate explanation requiring replication, not an established mechanism. DAM's score tracks the part of the mEPDS-9 that the PHQ-8 does not capture: partial $r$(DAM, mEPDS-9 $\mid$ PHQ-8) $= +0.279$ [0.11, 0.43], while partial $r$(DAM, PHQ-8 $\mid$ mEPDS-9) $= -0.179$ [$-0.33$, $-0.01$]. Item-level correlations locate that content precisely. DAM tracks affective distress; crying ($r=0.26$), sadness ($r=0.21$), feeling overwhelmed ($r=0.20$), anxiety ($r=0.19$), and is flat on the two positively-worded anhedonia items ($r\leq0.03$). On the PHQ-8 every item sits at or near zero except fatigue ($r=0.11$), with slightly negative associations for appetite change and anhedonia (both $-0.08$). This is not an anxiety effect: the EPDS anxiety subscale ($r=0.185$) and its depression subscale ($r=0.216$) correlate with DAM equally. The PHQ-8 contains substantial somatic and neurovegetative content (sleep, appetite, energy, psychomotor change) and these symptoms are common in pregnancy for reasons unrelated to mood. If that content contributes disproportionately to the PHQ-8 total in this cohort, a model that responds to affective distress would fail against the PHQ-8 here while succeeding against the same instrument in general-psychiatric adults, where those items do track depression.

\textit{Transfer also varies with elicitation task.} Within the prenatal cohort and against the same instrument, DAM reached 0.653 [0.550, 0.753] on spontaneous pregnancy narrative, 0.597 [0.486, 0.699] on open-ended opinion speech and 0.544 [0.434, 0.651] on the scripted reading passage; an ordering that follows similarity to its training distribution of spontaneous speech about personal experience.

\textit{Sex composition explains none of it.} DAM performed comparably on female and male general-psychiatric speech (0.657 vs 0.708 on DAIC-WOZ; 0.662 vs 0.710 on E-DAIC; $\Delta \approx -0.05$, $p\approx0.56$), yet dropped to 0.510 against the PHQ-8 in pregnant women. A model that handles women in one setting and fails on women in another is not failing because of the sex of the speakers. In the forward direction, the prenatal model performed better on female than male external participants (0.587 vs 0.461 on DAIC-WOZ; 0.588 vs 0.417 on E-DAIC), and a probe confirms why: out of domain, its score predicts participant sex (AUC 0.62--0.63) better than depression (0.48--0.51). Restricted to women it still did not separate from chance.

\textit{How large is the shift the models face?} A classifier trained to distinguish prenatal from external recordings, using the same 20 features the transfer model itself selects, reaches AUC 0.89 (DAIC-WOZ) and 0.88 (E-DAIC). Domain identity is thus far more recoverable from this feature space than depression is, which is why a model fitted in one domain carries so little into another, and why no component of the compound shift can be isolated from these data alone.

\begin{table}[H]
\centering
\caption{Transfer matrix (AUC, 95\% CI). Bold marks the three estimates whose confidence interval excludes chance. All cross-domain cells are pure inference, with no refitting, recalibration or threshold selection on the target cohort; the two prenatal cells for the prenatal-trained model are leakage-controlled cross-validated estimates within its own cohort, since a model cannot be applied to its own training cohort without being fitted on it. The two prenatal columns differ only in which self-report instrument defines the outcome; the participants, recordings, and models are identical. \textsuperscript{a}Prenatal-trained leakage-controlled model, acoustic features fused with community-health embeddings. Community-health embeddings do not exist for DAIC-WOZ or E-DAIC (no ZIP code), so this row has no external cells. \textsuperscript{b}The same prenatal-trained model using the acoustic block alone (Reading Passage, PHQ-8$\geq$10 training labels). This row is the like-for-like comparison with DAM, which is acoustic-only in every column, and is the only row that can be read across all four columns. \textsuperscript{c}DAM applied without refitting; prenatal values are aggregated per participant across tasks. The prenatal DAM set is N=136: participants with audio meeting DAM's minimum-duration requirement, after the same participant-grouping exclusions applied to the main analysis. It is not nested within the N=316 analytical sample, because DAM requires neither a ZIP-code match nor the reading-passage recording; robustness across three cohort definitions is reported in the text. Abbreviations: PDFM, Population Dynamics Foundation Model; DAM, Kintsugi Depression/Anxiety Model; mEPDS-9, modified nine-item Edinburgh Postnatal Depression Scale; DAIC-WOZ, Distress Analysis Interview Corpus (Wizard-of-Oz); E-DAIC, Extended DAIC.}
\label{tab:crossdomain}
\setlength{\tabcolsep}{3pt}\footnotesize
\begin{tabular}{lcccc}
\toprule
 & \multicolumn{2}{c}{\textbf{Prenatal cohort}} & & \\
\cmidrule(lr){2-3}
\textbf{Model} & \textbf{vs PHQ-8$\geq$10} & \textbf{vs mEPDS-9$\geq$10} & \textbf{DAIC-WOZ} & \textbf{E-DAIC} \\
\midrule
Prenatal, acoustic + PDFM\textsuperscript{a} & 0.494 [0.431, 0.560] & 0.456 [0.384, 0.525] & --- & --- \\
Prenatal, acoustic only\textsuperscript{b} & 0.452 [0.389, 0.511] & 0.474 [0.402, 0.545] & 0.505 [0.417, 0.593] & 0.478 [0.398, 0.564] \\
Kintsugi DAM\textsuperscript{c} & 0.510 [0.411, 0.609] & \textbf{0.645 [0.541, 0.744]} & \textbf{0.706 [0.620, 0.785]} & \textbf{0.708 [0.627, 0.789]} \\
\bottomrule
\end{tabular}
\end{table}

\section*{Discussion}

\subsection*{Principal Findings}

This study yields four findings. First, and centrally: whether a depression voice biomarker transfers into a new clinical setting is substantially shaped by the outcome instrument, not by the population alone. A general-population speech model applied without refitting to the same pregnant women, on the same recordings, reached 0.510 against the PHQ-8 and 0.645 against the modified EPDS-9. Nothing about the model, the speakers or the audio differed between those two numbers: only the questionnaire that defined depression. Second, exploratory item-level associations are consistent with a construct mismatch: the model's score was associated with selected affective-distress items and showed no association with the somatic PHQ-8 items examined, although only one of 17 item-level tests survived multiplicity adjustment. Third, transfer additionally varies with elicitation task, following similarity to the training distribution (pregnancy narrative $>$ open-ended speech $>$ scripted reading). Fourth, our own prenatal-trained model performed at chance both within the cohort, on the pre-registered outcome, and on two independent external corpora; no cell of a 16-cell pre-specified grid survived multiplicity adjustment under either of two modeling paradigms.

The practical consequence is that ``does this voice biomarker generalize?'' is not a well-posed question. It becomes well-posed only when the population, the elicitation task and the outcome instrument are named together. A tool validated against one instrument should not be assumed to work against another, even in the same people.

\subsection*{Neither Voice nor Community Context Discriminates Individuals}

The voice increment over community context is $\Delta\text{AUC}=-0.047$ at the pre-registered outcome, and community context itself does not discriminate either: the PDFM arm reaches a median of 0.519 across the 16 cells, a maximum of 0.546, and clears chance in none of them. Community context performs no worse than acoustics, but both are at chance, so neither carries individual-level signal.

The like-for-like comparison sharpens this. Restricted to the acoustic block alone, which is the only basis on which our model and the general-population comparator can be compared directly, the prenatal-trained model reaches 0.452 against the PHQ-8 and 0.474 against the mEPDS-9, while the comparator reaches 0.510 and 0.645 on the same recordings (Table~\ref{tab:crossdomain}). Both operate on the voice signal alone, without community context, transcripts or refitting. The difference between them is therefore not the information available in the audio but what each model has learned to extract from it, which is consistent with the null reported here being a property of this cohort's features and fitting procedure, and not evidence that perinatal speech carries no recoverable signal at all.

A graph-based foundation model pretrained on a large national survey showed the same absence of individual-level transferability in this cohort (Multimedia Appendix~10). This is an individual-level result and should not be read more broadly. Population-level associations between geospatial context and perinatal depression are well documented, including in work using these same embeddings at the level of aggregated geographies. A representation can track prevalence across places and still fail to rank individuals within them; that is the ordinary distinction between ecological and individual inference, and our result speaks only to the latter.

\subsection*{Implications for Validation Practice}

Our results argue for three changes in how voice-based mental-health tools are evaluated. First, validation claims should name the population, the elicitation task and the outcome instrument together; any one alone is insufficient, and our data show the instrument can flip a result from chance to above chance in the same participants. Second, a model deployed outside the setting it was validated in gives no signal that it has left that setting; DAM's score distribution looks the same whether it is discriminating at 0.71 or at 0.51, which argues for explicit detection of the mismatch, not for trust in the output. In our data the information needed for such detection is present: domain identity is recoverable at AUC 0.88 from the same 20 features the model uses. Third, the instrument should be chosen for what the model can plausibly detect. In pregnancy specifically, somatic-weighted instruments are poor targets for acoustic models because gestation elevates the somatic items independently of mood.

We do not report a deployment scenario for this cohort. With the pre-registered outcome at chance and a non-informative operating point (LR$+$ $=1.02$, LR$-$ $=0.96$), no configuration examined here supports screening use.

\subsection*{Cross-Domain Non-Transfer Is Robust to Audio Format}

The cross-domain failure is not an audio-format artifact: format-matching Voice-AI audio to DAIC's native 16 kHz PCM and re-running the comparator model leaves $\Delta\text{AUC}$ essentially centered on zero, and a symmetric reversal test does not rescue transfer in either direction. Browser-recorded audio in this cohort was, if anything, higher quality than the laboratory reference corpora. Full analyses, including the task-conditional partial-transfer pattern and the pregnancy-specific phonatory substrate, are in Multimedia Appendix 9.

\subsection*{Comparison With Prior Work}

The voice-biomarker literature has repeatedly reported above-chance, sometimes high, within-study discrimination for depression, but external validation across clinical settings is rare and controls for data leakage are inconsistently applied.\cite{low2020,koops2023,kappen2023,kapoor2023leakage} Our results add two correctives. First, generalization failure in this literature is routinely attributed to population differences; we show that in the one direction where the two can be separated, the outcome instrument determined whether transfer occurred, with the population held fixed. Reports of population-specific failure should therefore be read with the instrument in view, since studies comparing cohorts almost always change instruments at the same time. Second, our own within-cohort estimates collapse to chance under leakage-controlled evaluation with multiplicity adjustment, consistent with the view that some previously reported performance reflects analytic flexibility and leakage, not transportable signal.

\subsection*{Limitations}

(1)~\emph{Self-report labels at a single timepoint.} Both outcomes are self-administered screening instruments, not clinical diagnoses, collected once. This is the study's most consequential limitation and it is compounded by (2). (2)~\emph{The mEPDS-9 required item-level rescoring.} The instrument as administered used a generic response scale in place of the published EPDS anchors, and the responses required rescoring at the item level before analysis. Because the resulting totals are indistinguishable from valid ones at the score-total level, we recommend item-level verification of any self-report label before modeling. (3)~\emph{Cell-level instability.} Individual cells of the task $\times$ instrument grid vary by up to 0.08 across protocol variations, so no single cell should be interpreted; this is why we report the whole grid with multiplicity adjustment and interpret no single best-performing configuration. (4)~\emph{Post-hoc analyses.} The instrument-specificity finding, the item-level mechanism, the task ordering and the sex-stratified analyses were not pre-specified. They are hypothesis-generating and require pre-registered replication. (5)~\emph{The DAM subsample is a different sample.} DAM analyses use the N=136 participants whose recordings meet its minimum-duration requirement and who pass the same participant-grouping exclusions. This set is not nested within the N=316 analytical sample: DAM requires neither ZIP matching nor the reading-passage recording, so the two analyses are not drawn from identical participants. The instrument contrast is internally consistent because both estimates come from the same participants; the comparison against the within-cohort analysis is not exact. (6)~\emph{Compound shift.} Population, task and instrument differ simultaneously between cohorts; we can isolate the instrument only in the direction where two instruments exist on the same participants, and cannot isolate population from task at all. (7)~\emph{No depression-history data.} Onset timing was not collected, so pre-existing and incident antenatal depression cannot be distinguished. (8)~\emph{Residual demographic differences.} Sex is addressed directly, but age could not be matched to the prenatal range within the external corpora without collapsing sample size. (9)~\emph{Online recruitment selection.} Cohort prevalence (53--74\%) far exceeds general perinatal populations; PPV/NPV/NNS are recomputed at 12--30\% deployment prevalence (Table~\ref{tab:screening}). (10)~\emph{Self-harm-item omission} under the IRB determination. Additional limitations appear in Multimedia Appendix 7.

\subsection*{Scope of Inference}

The supported claims are narrow and we state them explicitly. Within this cohort, under leakage-controlled evaluation, neither acoustic features nor community-health context discriminated individuals on the pre-registered outcome, and no cell of the pre-specified grid survived multiplicity adjustment. Across cohorts, a general-population speech model transferred into pregnancy against one instrument and not another, with an item-level mechanism consistent with that pattern. We do not claim that voice biomarkers cannot work in perinatal populations; we claim that this cohort, with these instruments and these tasks, provides no evidence that they do, and that a substantial part of what looks like population-specific failure in this literature may be instrument-specific.

\subsection*{Future Directions}

Three directions follow directly. \emph{Modeling.} Parameter-efficient fine-tuning of foundation speech models on target-domain data is the most direct test of whether the failures we observe are trainable domain gaps; the task-conditional partial transfer we report suggests adaptable representations exist. Unsupervised domain adaptation is motivated by the size of the shift we measure, and our domain probe sets its difficulty. Domain-invariant objectives should target the axes our analysis identifies (population, task and instrument) and the sex-tracking probe suggests speaker-attribute invariance deserves an explicit penalty. \emph{Measurement.} Confirmatory work should adopt structured clinical interview as the primary outcome, and where self-report is used, should report item-level psychometrics instead of totals alone. In perinatal samples specifically, instruments should be chosen or scored so that pregnancy-confounded somatic content does not dominate the target. \emph{Design.} Studies should vary population, task and instrument separately rather than together; a cohort carrying two instruments on the same participants, as ours did, is what made the central finding of this paper visible at all.

\section*{Conclusions}

Whether a depression voice biomarker transfers to a new clinical setting may depend on what the outcome instrument measures, not on the population alone. A general-population speech model applied without refitting to the same pregnant women reached AUC 0.645 against a modified EPDS-9 and 0.510 against the PHQ-8 (identical participants, recordings and model) with exploratory item-level results consistent with the model responding to affective-distress content, not to the somatic PHQ-8 content that gestation itself elevates. Our own prenatal-trained model performed at chance within the cohort on its pre-registered outcome and on two independent external corpora, and no cell of a pre-specified 16-cell grid survived multiplicity adjustment. Validation of voice-based mental-health tools should therefore specify population, elicitation task and outcome instrument together, and should not assume that a tool validated against one instrument works against another in the same people.

\section*{Acknowledgments}

The teams of Cint, TELL and Chessboard Labs (Databoard) have gone above and beyond in supporting this project. We thank them. We also thank the Population Dynamics Foundation Model team at Google Research for sharing the model weights used in the community-health analyses.

\section*{Funding}

nurtur health, Inc partially funded data collection. Beyond the contributions of the authors affiliated with it, which are itemized under Authors' Contributions, the funder had no role in the design and conduct of the study; collection, management, analysis, and interpretation of the data; preparation, review, or approval of the manuscript; or decision to submit the manuscript for publication. Adolfo M.\ Garcia is partially supported by the National Institute On Aging of the National Institutes of Health (R01AG075775, R01AG083799, 2P01AG019724); ANID (FONDECYT Regular 1250317, 1250091); Agencia Nacional de Promocion Cientifica y Tecnologica (01-PICTE-2022-05-00103); and the Multi-partner Consortium to Expand Dementia Research in Latin America (ReDLat), which is supported by the Fogarty International Center and the National Institutes of Health, the National Institute on Aging (R01AG057234, R01AG075775, R01AG21051, and CARDS -- NIH), Alzheimer's Association (SG-20-725707), Rainwater Charitable Foundation's Tau Consortium, the Bluefield Project to Cure Frontotemporal Dementia, and the Global Brain Health Institute. Other authors report no specific funding for this project.

\section*{Conflicts of Interest}

MGF reports employment at Health Stream Analytics, LLC; MGF and SDF report employment at nurtur health, Inc. nurtur health, Inc partially funded the data collection for this study (see Funding). JS (Schwoebel) and JS (Shor) report employment at Quome Inc. MC and AMG report employment at TELL Inc. RLW reports no conflicts of interest. 

\section*{Declaration of Generative AI Use}

The authors used a generative-AI assistant (Anthropic Claude, accessed through the Claude Code interface, during 2025--2026) during the preparation of this work. Following the GAIDeT taxonomy, the following task categories were delegated to the tool under author direction and supervision:

\begin{itemize}\itemsep0pt
\item \textbf{Conceptualization}: limited to the revision. The study's conception, research questions, analysis plan and its pre-registration precede any use of generative AI. During revision, the tool proposed the reframing of the central claim from population-specific to instrument-specific transfer failure, which the authors evaluated against the data and adopted.
\item \textbf{Methodology}; limited to the revision. Selection of analysis methods including the multiplicity-adjustment procedure, bootstrap contrasts, participant-grouped cross-validation, the domain-separability and sex-attribution probes, and the partial-correlation decomposition. The original study design and protocol were author-determined without AI involvement.
\item \textbf{Software development and automation}: generation and refactoring of the feature-extraction, cross-validation and statistical-analysis code, including the leakage-control and negative-control harnesses.
\item \textbf{Data management}; data cleaning, statistical computation, and figure and table generation.
\item \textbf{Writing and editing}: drafting and copy-editing of manuscript sections, journal-format conversion, and preparation of the response to reviewers.
\item \textbf{Literature review}; assistance with searching for and locating relevant literature. Assessment of the retrieved sources, and all use made of them in the manuscript, are the authors' own.
\item \textbf{Supervision}: code and results checking. This surfaced the instrument-scoring error that required the item-level rescoring described under Limitations, an inconsistency between cohort definitions in the comparator analyses, and a unit error in the flow diagram, each of which the authors then verified independently.
\end{itemize}

Ethics review was not delegated. All interpretation of results, all decisions about what to report, and the final content of the manuscript are the authors' own. No AI system is listed as an author. Every AI-assisted output was reviewed and verified by the authors, who take full responsibility for the integrity and accuracy of this work, including any content produced with AI assistance. M.G.F. delegated the tasks listed above.

\section*{Authors' Contributions}

MGF and JS (Schwoebel) conceived and designed the study and acquired the data. MC and AMG developed the voice processing and acoustic--linguistic feature-extraction pipelines. MGF designed and implemented the leakage-controlled evaluation protocol, the negative-control harness, and the external-validation and community-health (PDFM) analyses. MGF and JS (Shor) performed the data analysis. RLW, SDF, and AMG provided clinical and psychometric interpretation. RLW and MGF drafted the manuscript. All authors critically revised the manuscript for important intellectual content and approved the final version.

\section*{Data Availability}

The prenatal voice dataset is not publicly available due to the terms of the IRB determination and participant consent, which did not authorize data sharing. The DAIC-WOZ and E-DAIC datasets are available through the University of Southern California's Institute for Creative Technologies upon execution of a data use agreement. The Kintsugi DAM model is open-source (Apache 2.0) and available at the Kintsugi Health GitHub repository. PDFM embeddings are publicly available from Google Research. Analysis code, including the instrument scoring and an independent verification script, together with machine-readable result files, are available from the corresponding author on reasonable request. They are not deposited publicly because the result files derived from this cohort contain participant-level outcome values that participant consent does not authorise us to share.

\section*{Abbreviations}

RID: respondent identifier, the participant identifier assigned by the recruitment panel.

\noindent
AUC: area under the receiver operating characteristic curve\\
BH: Benjamini--Hochberg\\
CI: confidence interval\\
CV: cross-validation\\
DAIC-WOZ: Distress Analysis Interview Corpus -- Wizard-of-Oz\\
DAM: Depression/Anxiety Model (Kintsugi Health)\\
E-DAIC: Extended Distress Analysis Interview Corpus\\
EPDS: Edinburgh Postnatal Depression Scale\\
mEPDS-9: modified nine-item Edinburgh Postnatal Depression Scale (item 10 omitted)\\
F0: fundamental frequency\\
HNR: harmonics-to-noise ratio\\
IPW: inverse probability weighting\\
IRB: institutional review board\\
LR: logistic regression\\
LR$+$/LR$-$: positive/negative likelihood ratio\\
MFCC: mel-frequency cepstral coefficient\\
NNS: number needed to screen\\
NPV: negative predictive value\\
OSF: Open Science Framework\\
PDFM: Population Dynamics Foundation Model\\
PHQ-8: eight-item Patient Health Questionnaire\\
PPV: positive predictive value\\
TPOT: Tree-based Pipeline Optimization Tool\\
ZCTA: ZIP Code Tabulation Area

\newpage

\appendix
\section*{Multimedia Appendices}
\sloppy

\renewcommand{\thesection}{e}
\renewcommand{\thesubsection}{e\arabic{subsection}}
\renewcommand{\thetable}{S\arabic{table}}
\setcounter{table}{0}

\subsection*{Multimedia Appendix 1: Leakage-Controlled Evaluation Protocol}

The main-text Results were generated under a leakage-controlled evaluation protocol designed to eliminate three sources of test-fold information leakage common in the voice-biomarker literature: (i) feature selection performed on the full cohort before cross-validation, (ii) classifier selection from a multi-classifier grid by accuracy on the test fold itself, and (iii) class balancing applied before the train/test split. We address all three by structuring the entire feature-selection-plus-classifier pipeline as an \texttt{sklearn} \texttt{Pipeline} passed to \texttt{cross\_val\_predict}, fixing the classifier a priori (or selecting it inside an inner cross-validation loop on training-fold data only), and replacing pre-split balancing with class-weighted classifiers. A 20-permutation negative-control harness validates each pipeline configuration before its real-data result is read; shuffled-label, random-feature, and joint-shuffle controls are required to sit at chance ($\approx$0.50 AUC) for the reported real-data result to be considered admissible. The pipeline scripts, the negative-control harness, and per-(task $\times$ threshold) result JSONs are available from the corresponding author on reasonable request (see Data Availability). We offer this protocol as a methodological template for the wider voice-biomarker literature, where leakage of these kinds is documented to inflate reported performance \cite{kapoor2023leakage}.

\subsection*{Multimedia Appendix 2: Study Design and Voice Recording Protocol}

Table~\ref{tab:studydesign} details the study protocol, including the sequence of vocal tasks, survey instruments, and mood assessments administered to participants. Three free (unstructured) speech recordings were made, first for process testing (task \#1), followed by two recordings (task \#2 and \#4) on pregnancy-related or unrelated contexts. A structured self-affirmation voice recording was also made (task \#3), placed deliberately in the order of recordings to minimize affective spill-over between the two free speech recordings. Finally, PHQ-8 and modified EPDS were administered, separated by five basic survey questions on the opinions and perceptions related to perinatal health care.

{\small
\begin{longtable}{p{3cm} p{1.7cm} p{1.3cm} p{2.2cm} p{5.1cm}}
\caption{Study design: sequence of tasks and instruments administered to participants.}\label{tab:studydesign}\\
\toprule
\textbf{Task} & \textbf{Duration} & \textbf{Voice Sample} & \textbf{Category} & \textbf{Prompt / Question} \\
\midrule
\endfirsthead
\multicolumn{5}{l}{\textit{Table S1 (continued)}}\\
\toprule
\textbf{Task} & \textbf{Duration} & \textbf{Voice Sample} & \textbf{Category} & \textbf{Prompt / Question} \\
\midrule
\endhead
\bottomrule
\endlastfoot
Consent page & --- & No & Administrative & Study purpose, risks, procedures, privacy \\
\addlinespace
Microphone check & --- & No & Setup & Level check (``detecting voice'') ensures a quiet environment \\
\addlinespace
Free speech 1 (Sound Check) & 10 seconds & Yes & Adapted from Voiceome & ``Please say, `the quick brown fox jumps over the lazy dog.' This is a sample question to make sure you can record your audio on our platform.'' \\
\addlinespace
Free speech 2 (Pregnancy Narrative) & 1 minute & Yes & Pregnancy-related & ``Please share how you have experienced your pregnancy over the past month, focusing on your emotions and feelings. Kindly provide as many details as possible.'' \\
\addlinespace
Visual sliding mood scale & 10 seconds & No & Likert -- Mood & ``On a scale of 0--100, how sad are you (100 being most sad, 0 not being sad)?'' \\
\addlinespace
Self-affirmation (Reading Passage) & 2 minutes & Yes & Resilience & Read-aloud passage: ``My body knows how to bring life and grow\ldots{} Peace flows through me like a quiet stream, nourishing my body and my spirit.'' \\
\addlinespace
Visual sliding mood scale & 15 seconds & No & Likert -- Mood & ``On a scale of 0--100, how sad are you (100 being most sad, 0 not being sad)?'' \\
\addlinespace
Free speech 3 (Free Speech) & 1 minute & Yes & Non-pregnancy related & ``Some people claim that the world has never been better than it is now, while others propose that we, as societies, become worse year after year. What is your opinion? Please state it clearly and provide as many reasons and examples as possible.'' \\
\addlinespace
Modified EPDS & 2 minutes & No & Survey & 9-item Edinburgh Postnatal Depression Scale (item 10, suicidality, removed) \\
\addlinespace
Survey questions & 90 seconds & No & Survey & Interest in perinatal mental health care solutions \\
\addlinespace
PHQ-8 & 2 minutes & No & Survey & 8-item Patient Health Questionnaire (item 9, suicidality, removed from PHQ-9) \\
\end{longtable}
}

\noindent\textit{Note:} Tasks 2 and 4 (Pregnancy Narrative and Free Speech) were presented in A/B counterbalanced order across two participant groups to control for order and fatigue effects. The pregnancy-related and non-pregnancy-related free speech prompts were swapped in 50\% of participants.

\subsection*{Multimedia Appendix 3: Feature Extraction Pipelines}

\textbf{Pipeline A} (Allie/Databoard) extracts low-level acoustic features that capture physical properties of speech production (spectral shape, vocal tract resonances, and energy distribution) without any semantic content.\cite{schwoebel2021longitudinal} The acoustic feature set comprises 325 features: 13 MFCCs, 26 log Mel-filterbank energies, and 26 spectral subband centroids, each with 5 summary statistics per recording. Additional descriptors include temporal features (articulation rate, pause duration, voiced segment length), prosodic features (F0 contour and variability, pitch range, intensity envelope), spectral parameters (energy across frequency bands, spectral centroid, flux, entropy), and voice quality measures (jitter, shimmer, harmonics-to-noise ratio). Because these features measure \textit{how} speech is produced rather than \textit{what} is said, Pipeline A enables screening that requires no speech transcription.

\textbf{Pipeline B} (TELL) provides the linguistic and conversational interaction dimension through a validated clinical speech analysis framework,\cite{garcia2024tell,garcia2025tell,ulkumen2022,pereztoro2025,ferrante2024,eyigoz2020,sanz} extracting higher-order features including verbosity, sentiment, semantic granularity, and conversational timing. Speech preprocessing included spectral-gating noise reduction, diarization using MarbleNet VAD and TitaNet speaker embeddings, and transcription via Whisper Large-v3.

\paragraph*{Features documented but \emph{not} used in the headline analysis.} Two additional feature sources are documented in our pipeline code but are \emph{deliberately excluded from the main-text headline analysis}, the calibration grid, and the TPOT-CV sensitivity analysis. We disclose this explicitly so the headline feature set is unambiguous:

\begin{itemize}\setlength{\itemsep}{0pt}
\item \textbf{Pipeline B (TELL) linguistic features, \emph{excluded}.} The TELL framework ($\sim$130 features: verbosity, sentiment, semantic granularity, conversational timing) is documented above for transparency but excluded from the main-text analysis for two converging reasons: (i) it requires speech-to-text transcription, which adds infrastructure, latency, and a semantic-content privacy surface that the privacy tier explicitly avoids; (ii) under the leakage-controlled sensitivity analysis, adding semantic transcript features did \emph{not} improve AUC and on Free Speech actively reduced it ($\Delta\text{AUC}=-0.016$, DeLong $p=0.055$), most likely a feature-count-inflation penalty at the pre-filter stage.
\item \textbf{Kintsugi Depression/Anxiety Model (DAM) scores, \emph{external comparator}.} The DAM model (a Whisper-fine-tuned deep-learning depression/anxiety scorer; details in Multimedia Appendix~4) is applied by pure inference, without refitting, for (a) benchmarking on DAIC-WOZ and E-DAIC, (b) the within-prenatal instrument contrast reported in the main text (N=136), and (c) a three-way (acoustic + DAM + community context) sensitivity analysis on the N=115 three-way-complete subset, where it provides no incremental contribution. DAM is \emph{not} part of the prenatal acoustic + community-context model whose within-cohort performance is reported as the pre-registered outcome.
\end{itemize}

The headline analysis is therefore the 2-way late fusion of the 375-feature privacy-tier audio block with the 330-dimensional PDFM embedding.

\subsection*{Multimedia Appendix 4: Kintsugi DAM Model Details}

The Kintsugi Health Depression/Anxiety Model (DAM) is an open-source model (Apache 2.0 license) that fine-tunes OpenAI's Whisper-large-v2 speech encoder using Low-Rank Adaptation (LoRA).\cite{kintsugi_dam,hu2021lora,radford2023} The model was trained on 863 hours of telephone calls from approximately 35,000 individuals with PHQ-9 labels. It outputs continuous depression and anxiety scores and quantized severity levels (0--3).

For our prenatal cohort, audio files were converted from AAC-in-WAV container at 48 kHz to PCM float32 at 16 kHz. Of 1,128 recordings (4 tasks $\times$ 283 participants with DAM-processable audio), 422 files from 149 participants met the minimum 30-second duration requirement across three eligible tasks. DAM scores were aggregated per participant by averaging across tasks.

For E-DAIC, audio was directly loaded at its native 16 kHz format. All 219 participants with PHQ-8 labels were processed. DAM's mean score rises monotonically across PHQ-8 severity bands on that corpus (Table~\ref{tab:dam_severity}), confirming that the model behaves as intended within its own domain.

\begin{table}[H]
\centering
\caption{Mean DAM depression score by PHQ-8 severity band on E-DAIC. The score is the model's continuous output, not a discrimination metric; it rises monotonically across bands.}
\label{tab:dam_severity}
\begin{tabular}{lcc}
\toprule
\textbf{PHQ-8 Range} & \textbf{N} & \textbf{DAM Depression (mean $\pm$ SD)} \\
\midrule
None (0--4) & 86 & $-$1.043 $\pm$ 0.314 \\
Mild (5--9) & 46 & $-$0.920 $\pm$ 0.324 \\
Moderate (10--14) & 30 & $-$0.842 $\pm$ 0.347 \\
Mod-severe (15--19) & 20 & $-$0.655 $\pm$ 0.314 \\
Severe (20+) & 7 & $-$0.612 $\pm$ 0.376 \\
\bottomrule
\end{tabular}
\end{table}

DAM scores show a clear monotonic relationship with PHQ-8 severity on E-DAIC, confirming the model captures depression-relevant acoustic information within its target domain.

\subsection*{Multimedia Appendix 5: Bias-Aware Deployment Assessment}

\paragraph*{Sociodemographic characterization of the respondent-identifier (RID)-linked cohort.} Table~\ref{tab:s_sociodemo} summarizes the self-reported sociodemographic distribution of the RID-linked cohort (N=390, before the leakage-controlled exclusions that define the N=316 analytical sample), stratified by mEPDS-9$\geq$10. Covariates are complete for all but a small number of participants, who appear in the ``Prefer not to answer'' category where the panel offered one. Distributions do not differ significantly between mEPDS-9$\geq$10-positive and -negative groups for age, education, employment, race or household income. Two covariates do differ at the nominal level: relationship status ($p=0.030$) and smoking status ($p=0.001$; see footnote). Neither is adjusted for the six covariates examined, and both describe the recruited sample and do not provide evidence about the outcome.

{\small
\begin{longtable}{p{2.3cm} p{4.6cm} p{2.05cm} p{2.05cm} r}
\caption{Sociodemographic characteristics of the RID-linked cohort ($N=387$ unique participants with a scorable mEPDS-9, shown before the leakage-controlled exclusions that define the $N=316$ analytical sample), stratified by mEPDS-9$\geq$10. The table describes the recruited cohort and is not used to support any performance claim. Counts (column \%). $p$-values from Mann-Whitney U for age and Chi-square for nominal categories, unadjusted for the six covariates examined. Response categories are assigned from the recruitment panel's own codebook; every observed code maps to a displayed category, so no participant falls into an unclassified group.}\label{tab:s_sociodemo}\\
\toprule
\textbf{Variable} & \textbf{Category} & \textbf{mEPDS-9$\geq$10 (n=279)} & \textbf{mEPDS-9$<$10 (n=108)} & \textbf{$p$} \\
\midrule
\endfirsthead
\multicolumn{5}{l}{\textit{Table S3 (continued)}}\\
\toprule
\textbf{Variable} & \textbf{Category} & \textbf{mEPDS-9$\geq$10 (n=279)} & \textbf{mEPDS-9$<$10 (n=108)} & \textbf{$p$} \\
\midrule
\endhead
\bottomrule
\endlastfoot
Age (years, mean $\pm$ SD) & --- & 30.0 $\pm$ 7.1 & 30.5 $\pm$ 6.7 & 0.440 \\
\midrule
Education & High school or less & 77 (28\%) & 29 (27\%) & 0.276 \\
 & Some college / vocational / Associate & 96 (34\%) & 29 (27\%) &  \\
 & Bachelor's & 74 (27\%) & 31 (29\%) &  \\
 & Master's, professional, or doctorate & 31 (11\%) & 17 (16\%) &  \\
 & None of the above & 1 (0\%) & 2 (2\%) &  \\
\midrule
Employment & Employed full-time & 182 (65\%) & 76 (70\%) & 0.798 \\
 & Employed part-time & 32 (11\%) & 10 (9\%) &  \\
 & Student / homemaker / military & 27 (10\%) & 11 (10\%) &  \\
 & Not working / unable to work / other & 37 (13\%) & 11 (10\%) &  \\
 & Prefer not to answer & 1 (0\%) & 0 (0\%) &  \\
\midrule
Race & White & 159 (57\%) & 61 (56\%) & 0.920 \\
 & Black or African American & 83 (30\%) & 32 (30\%) &  \\
 & Asian & 11 (4\%) & 3 (3\%) &  \\
 & American Indian / Alaska Native / Pacific Islander & 12 (4\%) & 7 (6\%) &  \\
 & Some other race & 13 (5\%) & 5 (5\%) &  \\
 & Prefer not to answer & 1 (0\%) & 0 (0\%) &  \\
\midrule
Household income & $<$\$25{,}000 & 64 (23\%) & 20 (19\%) & 0.226 \\
 & \$25{,}000--\$49{,}999 & 76 (27\%) & 33 (31\%) &  \\
 & \$50{,}000--\$99{,}999 & 93 (33\%) & 29 (27\%) &  \\
 & \$100{,}000--\$199{,}999 & 38 (14\%) & 18 (17\%) &  \\
 & $\geq$\$200{,}000 & 7 (3\%) & 6 (6\%) &  \\
 & Prefer not to answer & 1 (0\%) & 2 (2\%) &  \\
\midrule
Relationship status & Married & 106 (38\%) & 59 (55\%) & \textbf{0.030} \\
 & Domestic partnership / engaged & 61 (22\%) & 15 (14\%) &  \\
 & Single, never married & 99 (35\%) & 32 (30\%) &  \\
 & Separated / divorced / widowed & 11 (4\%) & 1 (1\%) &  \\
 & Prefer not to answer & 2 (1\%) & 1 (1\%) &  \\
\midrule
Smoking status & Smokes & 187 (67\%) & 54 (50\%) & \textbf{0.001} \\
 & Does not smoke; uses other tobacco or vape & 41 (15\%) & 16 (15\%) &  \\
 & Does not smoke & 51 (18\%) & 38 (35\%) &  \\
\end{longtable}
}

\noindent\smallskip\footnotesize $^*$Two covariates are associated with mEPDS-9$\geq$10 at the nominal level: smoking ($p=0.001$) and relationship status ($p=0.030$); neither is adjusted for the six covariates tested. Inverse-probability weighting on the full covariate set changes discrimination by $\Delta\text{AUC}=-0.008$ on the leakage-controlled pipeline.\normalsize

\paragraph*{Confounding assessment and IPW.} Potential sociodemographic confounding was assessed through: (1) demographic group comparisons using Mann-Whitney U and Chi-square tests with effect sizes (Cohen's $d$, Cramer's $V$); (2) Inverse Probability Weighting (IPW) for the acoustic pipeline where significant demographic-depression associations were identified: propensity scores estimated via logistic regression with weights truncated to [0.1, 10.0] and normalized; and (3) voice-only model performance comparison.

Two observations bear on the equitable deployment of any future voice-based tool under this protocol. (i) Inverse-probability weighting on the full sociodemographic covariate set leaves discrimination essentially unchanged ($\Delta\text{AUC}=-0.008$ at the pre-registered outcome), consistent with the main-text finding that sociodemographic variables add no detectable value beyond privacy + PDFM. (ii) Community context does not separate the outcome groups in this cohort: the community-health arm's out-of-fold score differs between screen-positive and screen-negative participants by Cohen's $d=+0.01$ at the pre-registered outcome (arm AUC 0.508). There is therefore no geographic signal in these data for a deployed model to exploit, and no geography-driven differential performance to adjudicate. This is a statement about individual-level discrimination in this cohort and not about population-level associations between place and perinatal depression, which are well documented.

\subsection*{Multimedia Appendix 6: Feature Extraction Technical Validation}

Cross-toolkit comparison on 219 E-DAIC participants confirmed that our feature extraction captures established acoustic dimensions. Fundamental frequency (F0) showed near-perfect agreement ($r$=0.934), MFCC coefficient 1 showed strong agreement ($r$=0.865), and voice quality measures showed moderate agreement (shimmer $r$=0.639; energy $r$=0.607), consistent with expected differences in implementation-specific processing choices.

These toolkit comparisons are indicative rather than definitive: they were run under a separate protocol, so their absolute values are not directly comparable with the main-text estimates, and they are reported to establish only that the failure to transfer is not toolkit-specific. Feature importance was assessed using gradient boosting (5-fold CV, PHQ-8$\geq$10) with four feature sets: our pipeline (375 features, AUC=0.541), OpenSMILE eGeMAPS (115 features, AUC=0.483), OpenSMILE MFCCs (195 features, AUC=0.546), and combined (462 features, AUC=0.563). All feature sets performed near chance on the general-psychiatric E-DAIC corpus, indicating that the failure to transfer is not an artifact of any one feature-extraction toolkit. It does not establish the cause of that failure, which the design confounds across population, task, recording channel and outcome instrument.

\subsection*{Multimedia Appendix 7: Additional Limitations}

Additional methodological details on limitations: (1) The 10-second Sound Check yielded chance-level performance on the analytical sample (AUC 0.457--0.550 across the four outcome definitions; Table~\ref{tab:multiplicity} and Multimedia Appendix~8), indicating insufficient duration for the feature-extraction methods used here. (2) Multimodal fusion with TELL linguistic features requires access to multiple data sources; the simpler acoustic + PDFM combination reported here does not require TELL infrastructure. (3) \textbf{Feature-to-sample ratio.} Within each cross-validation training fold, \texttt{SelectKBest} reduced 375 acoustic features and 330 PDFM dimensions to 20 each before classifier fitting, yielding approximately 16 samples per feature on the analytical sample (N=316; 20 features for the per-task N range 314--322 across the four speech tasks). (4) \textbf{Cohort differences.} The prenatal cohort differs from both external corpora on sample size, recruitment channel, recording setting and outcome instrument simultaneously (Table~\ref{tab:crossdataset}), which is why the cross-domain comparison is a compound shift.

\begin{table}[H]
\centering
\caption{Cross-dataset comparison. \textsuperscript{*}Prevalence is computed on the same rows each analysis uses: the $N=316$ analytical sample for the prenatal cohort, and all labelled participants for the two external corpora ($N=189$ and $N=219$), in every case at a PHQ-8 total of 10 or above.}
\label{tab:crossdataset}
\setlength{\tabcolsep}{4pt}\small
\begin{tabular}{llll}
\toprule
\textbf{Dimension} & \textbf{Our Cohort} & \textbf{DAIC-WOZ} & \textbf{E-DAIC} \\
\midrule
N & 316 & 189 & 219 (labeled) \\
Population & Prenatal women & General psychiatric & General psychiatric \\
Depression prevalence (PHQ-8$\geq$10)\textsuperscript{*} & 55.8\% & 30.2\% & 29.7\% \\
Audio task & Structured (1--2 min) & Clinical interview & Clinical interview \\
Gender & 100\% Female & Mixed & Mixed \\
Recording setting & Remote (browser) & In-person lab & In-person lab \\
Audio quality & Variable & Controlled lab & Controlled lab \\
\bottomrule
\end{tabular}
\end{table}

\subsection*{Multimedia Appendix 8: AutoML Sensitivity Analysis}

\textit{Purpose.} This appendix documents the AutoML (TPOT-CV) sensitivity analysis, which triangulates the main-text conclusions under an automated pipeline-selection paradigm rather than a fixed classifier.

\textit{Protocol.} Outer 5-fold cross-validation with participant-grouped folds (\texttt{StratifiedGroupKFold} on participant identifier, with an assertion per fold that no participant and no row index appears in both training and test partitions). On each outer-training partition, \texttt{StandardScaler} and \texttt{SelectKBest}(k=20) are fitted on training rows only, and TPOT runs its own internal cross-validation on those rows to select a pipeline; the outer-test partition is untouched until prediction. Arms are fused by the unweighted mean of their out-of-fold probabilities, so no fusion weight is fitted. Search budget 3 minutes per outer fold, \texttt{search\_space=\textquotesingle linear-light\textquotesingle}. Total compute 17.3 wall hours. Per-fold training and test index hashes are retained for audit.

\textit{Result.} Across the 16 task $\times$ outcome cells the maximum AUC was 0.569 and one cell reached the nominal 5\% level ($p=0.020$); no cell survived Benjamini--Hochberg adjustment. The consolidated comparison across both modelling paradigms is Table~\ref{tab:multiplicity} in the main text.

\textit{Cell-level estimates are not stable.} Repeating the search under protocol variations reproduces individual cell values only loosely (mean absolute difference 0.036, maximum 0.081, correlation across the 16 cells $r=0.376$). The search is stochastic under a wall-clock budget, and values shift further with the fold-grouping scheme and with whether item-complete subsets are used. Individual cells in this grid should therefore not be interpreted, and selecting a primary outcome by choosing the best-performing cell would be reading noise.

\textit{Three-way sensitivity.} On the subset with acoustic, comparator-model and community-context features jointly available (N=115), no block among seven feature combinations reached nominal significance (maximum AUC 0.576), and adding the comparator model to voice + community context produced no increment in any cell (largest $\Delta$AUC $=+0.034$, 95\% CI $-0.012$ to $+0.083$).

\subsection*{Multimedia Appendix 9: Audio Format, Task Content, and the Limits of the Cross-Domain Comparison}

\textit{The cross-domain failure is not an audio-format artifact.} Format-matching the prenatal audio to the external corpora's native 16 kHz PCM and re-running the comparator model leaves $\Delta$AUC essentially centred on zero, and a symmetric reversal test does not rescue transfer in either direction. Characterisation of the recordings indicates that the browser-collected audio in this cohort was, if anything, of higher technical quality than the laboratory reference corpora, so the failure is not explained by degraded input.

\textit{Task content matters.} Within the prenatal cohort and against the same instrument, the comparator model performs best on spontaneous pregnancy narrative, less well on open-ended opinion speech, and least well on the scripted reading passage (main text). This ordering follows similarity to the model's training distribution of spontaneous speech about personal experience. It is descriptive: these cells were not compared formally, and the ordering should not be read as three distinguishable levels.

\textit{Pregnancy-specific phonatory substrate.} One candidate contributor to cross-domain difficulty is worth stating explicitly. Depression and pregnancy both alter phonation, through partly distinct mechanisms: psychomotor retardation and altered laryngeal-muscle tone for depression; progesterone-mediated mucosal change and altered respiratory mechanics for pregnancy.\cite{hancock2014,ghaemi2018,rechenberg2022} A model trained on general-population speech learns one set of substrate effects and a prenatal model another. This is a plausible mechanism, not one this study tests.

\textit{Units.} Where sample sizes appear in this appendix and its source analyses, counts of audio files and recordings are distinct from counts of sessions and participants: each participant contributed up to four recordings in one session. Analyses at the recording level are labelled as such and are not participant-level evidence.

\subsection*{Multimedia Appendix 10: Foundation-GNN (PRAMS-GAT) Transferability}

As a further sensitivity analysis, we evaluated whether a 4-layer graph-attention network pretrained on 1.05 million PRAMS records (PRAMS-GAT; 128-dimensional embeddings), previously applied in a single-site obstetric cohort with rich structured + ICD-10 features, could contribute orthogonal signal here. Only 4 of 312 PRAMS graph nodes (1.3\%) could be populated from the available Voice-AI demographic schema, versus approximately 50\% in the cohort where PRAMS-GAT was previously validated. The graph's Hispanic-origin node is among those that cannot be populated: the recruitment panel administered a single race item with no Hispanic/Latino level and no separate Hispanic-origin question, so the variable was never measured in this cohort. The resulting embeddings collapsed to rank-3 effective dimensionality, with median pairwise cosine distance 0.004 and 10\% of patient pairs producing identical embeddings. We therefore report PRAMS-GAT as not evaluable on the Voice-AI variable schema, and not as a comparator. This is a methodological observation: foundation-GNN transfer is gated by input feature density at the receiving cohort. A larger Stage 2 study collecting structured medical-history items or EHR-linked ICD-10 codes would render PRAMS-GAT viable as a comparator.

\end{document}

%% file: figure_consort.tex
\begin{figure}[htbp]
\centering
\begin{tikzpicture}[
  every node/.style={font=\small},
  bigbox/.style={draw, rectangle, minimum width=9.4cm, text width=9.0cm,
                 minimum height=11mm, align=center, thick, inner sep=1.6mm},
  excl/.style={draw, rectangle, minimum width=5.4cm, text width=5.0cm,
               minimum height=11mm, align=left, font=\small\itshape, dashed,
               inner sep=1.6mm},
  arrow/.style={-{Latex[length=2.5mm]}, thick}
]

\node[bigbox] (n1) {Voice + survey records assembled\\
  \textbf{N = 446 audio sessions}\\
  {\footnotesize 421 linked to 390 participants, 25 without a participant ID;
   31 participants contributed 2 sessions}};

\node[bigbox, below=9mm of n1] (n3) {Reading-Passage audio \emph{and} both\\
  depression scores present\\
  \textbf{N = 424 sessions}};
\node[excl, right=6mm of n3] (e3) {\textbf{--22 sessions}\\
  no Reading-Passage audio, or an incomplete depression score};
\draw[arrow] (n1) -- (n3);
\draw[arrow] ($(n1.south)+(0,-3mm)$) -| (e3.north);

\node[bigbox, below=9mm of n3] (n4) {PDFM-matched sample:\\
  valid ZCTA embedding by ZIP lookup\\
  \textbf{N = 377 sessions}};
\node[excl, right=6mm of n4] (e4) {\textbf{--47 sessions}\\
  ZIP absent from the PDFM index (rural / unmapped ZCTAs)};
\draw[arrow] (n3) -- (n4);
\draw[arrow] ($(n3.south)+(0,-3mm)$) -| (e4.north);

\node[bigbox, below=9mm of n4, very thick] (n4b) {\textbf{Leakage-controlled\\
  analytical sample}\\
  \textbf{N = 316 sessions}};
\node[excl, right=6mm of n4b] (e4b) {\textbf{--61 sessions}\\
  dual-session-artifact participants (both sessions) and audio-incomplete sessions};
\draw[arrow] (n4) -- (n4b);
\draw[arrow] ($(n4.south)+(0,-3mm)$) -| (e4b.north);

\node[bigbox, below=9mm of n4b, dashed] (n5) {Comparator-model (DAM) analysis set\\
  {\footnotesize any task with $\geq$30 s of audio; no ZIP match or reading passage required}\\
  \textbf{N = 136 sessions}};
\node[excl, right=6mm of n5] (e5) {\textbf{parallel branch}\\
  same participant-grouping exclusions, different audio requirement; 115 of these are also in the N = 316 sample};
\draw[arrow, dashed] (n4b) -- (n5);
\draw[arrow, dashed] ($(n4b.south)+(0,-3mm)$) -| (e5.north);

\end{tikzpicture}
\caption[CONSORT flow diagram of cohort assembly]{%
\textbf{CONSORT-style flow diagram of cohort assembly for the primary
(Reading-Passage, PHQ-8 $\geq$ 10) analysis.}
The participant-linkage line in the top box is
descriptive, not a filtering step. Exclusions are shown at the right of each step
and are mutually exclusive, so the counts are additive down the chain
($446 - 22 - 47 - 61 = 316$). Within the analytical sample, 312 sessions have a
complete item-level PHQ-8 and 313 a complete item-level modified EPDS-9; the
small shortfall is reported with each analysis rather than applied as a further
cohort filter. The dashed box is the comparator-model analysis set, a parallel
branch rather than a nested subset: that model requires at least 30 seconds of
audio from any eligible task but needs neither a ZIP-code match nor the
reading-passage recording, so the same participant-grouping exclusions are
applied to a different audio requirement; 115 of its 136 sessions also appear in
the $N=316$ sample. Per-task sample sizes for the secondary analyses are in
Multimedia Appendix~7.}
\label{fig:consort}
\end{figure}